\documentstyle [12pt,epsf]{article}
\begin{document}

\begin{center}
{\large \bf Theory of spin-polarized transport in ferromagnet-semiconductor
structures:\ Unified description of ballistic and diffusive transport}\\[2.0cm]
{R.\ Lipperheide, U.\ Wille\footnote{Corresponding author.  Phone:\
++49-30-80622685; FAX:\ ++49-30-80622098; E-mail:\ wille@hmi.de.}\\{\em
Abteilung Theoretische Physik, Hahn-Meitner-Institut Berlin,\\[-0.2cm]
Glienicker Str.\ 100, D-14109 Berlin, Germany}\\[2.0cm]}
\end{center}

\noindent
{\bf Abstract}\\[-0.6cm]

A theory of spin-polarized electron transport in ferromagnet-semi\-con\-ductor
heterostructures, based on a unified semiclassical description of ballistic and
diffusive transport in semiconductors, is outlined.  The aim is to provide a
framework for studying the interplay of spin relaxation and transport mechanism
in spintronic devices.  Transport inside the (nondegenerate) semiconductor is
described in terms of a thermoballistic current, in which electrons move
ballistically in the electric field arising from internal and external
electrostatic potentials, and are thermalized at randomly distributed
equilibration points.  Spin relaxation is allowed to take place during the
ballistic motion.  For arbitrary potential profile and arbitrary values of the
momentum and spin relaxation lengths, an integral equation for a spin transport
function determining the spin polarization in the semiconductor is derived.
For field-driven transport in a homogeneous semiconductor, the integral
equation can be converted into a second-order differential equation that
generalizes the spin drift-diffusion equation.  The spin polarization in
ferromagnet-semiconductor structures is obtained by matching the spin-resolved
chemical potentials at the interfaces, with allowance for spin-selective
interface resistances.  Illustrative examples are considered.  \\[0.0cm]


\newpage

\noindent
{\bf 1.\ Introduction}\\[-0.6cm]

In spintronics research, particular emphasis is currently placed on the study
of spin-polarized electron transport in heterostructures formed of a
nonmagnetic semiconductor and two (metallic or semiconducting) ferromagnetic
contacts \cite{gru02,aws02,sch02,zut04}.  For the actual design of spintronic
devices, a detailed theoretical understanding of this kind of transport problem
is indispensable.  Several pertinent studies have been performed so far, which
mostly rely on the drift-diffusion model.  A number of important results have
emerged.  (i) For an interface between a metallic ferromagnet and a
semi\-conductor without spin-selective interface resistance, the injected
current spin polarization is predicted to be very low owing to the large
conductivity mismatch \cite{sch00}.  (ii) Spin-selective interface resistances
arising from tunnel barriers or Schottky barriers can greatly enhance the
injection efficiency \cite{fil00,ras00,smi01,fer01,ras02a,alb03}.  (iii) A
similar effect is to be expected when a sufficiently high electric field is
applied across the semiconductor \cite{yuf02a}.  (iv) Under conditions where,
in the semiconductor, ballistic transport prevails over drift-diffusion, the
injection efficiency is controlled by the Sharvin interface resistance
\cite{kra03} unless spin-selective interface resistances are introduced.

While the theory of spin-polarized electron transport in
ferromagnet-semiconductor heterostructures has reached a level of considerable
sophistication, it appears that certain aspects of the semiconductor part of
the transport problem require a more systematic, unified treatment, such as the
interplay of spin relaxation and transport mechanism all the way from the
diffusive to the ballistic regime, and the effect of the detailed shape of the
electrostatic potential profile.  In this paper, we outline the principal ideas
of a theory that meets these requirements.  For illustrative purposes, we
present calculated results for the position dependence of the zero-bias current
spin polarization along a heterostructure as well as for the injected
polarization as a function of bias.  A detailed account of the formal
development as well as specific applications of our theory will be published
elsewhere \cite{lip05}.\\

\noindent
{\bf 2.\ Thermoballistic current}\\[-0.6cm]

Our treatment of spin-polarized electron transport relies on our previously
formulated unified semiclassical description of (spinless) electron transport
in parallel-plane semiconductor structures \cite{lip03}.  The description
is based on the concept of a ``thermoballistic electron current'' which
combines elements of the ballistic and diffusive transport mechanisms.  Here,
we briefly summarize this concept.

Assuming a one-dimensional geometry, we consider a (nondegenerate)
semiconducting sample enclosed between two plane-parallel ferromagnetic
contacts at $x = x_{1}$ and $x = x_{2}$, respectively, so that $S = x_{2} -
x_{1}$ is the sample length (see Fig.\ 1).  The electron current density
$J(x',x'')$ across the ``ballistic interval'' $[x',x'']$ between two
equilibration points $x'$ and $x''$ is given by
\begin{equation}
J(x',x'') = v_{e} N_{c} \, e^{-\beta E_{c}^{m}(x',x'')} \; \left[ e^{\beta
\mu(x')} - e^{\beta \mu(x'')} \right]
\label{eq:1}
\end{equation}
($x_{1} \leq x' < x'' \leq x_{2}$), which is the difference of the ballistic
current injected into the interval at its left end at $x'$ and the analogous
current injected at its right end at $x''$.  The function $\mu(x)$ is the
chemical potential at the equilibration point $x$.  Furthermore, $v_{e} = (2\pi
m^{*} \beta)^{-1/2}$ is the emission velocity, $N_{c} = 2(2\pi m^{*}/\beta
h^{2})^{3/2}$ is the effective density of states at the conduction band edge,
$m^{*}$ is the effective mass of the electrons, and $\beta = (k_{B} T)^{-1}$.
The current (\ref{eq:1}) contains only the transmitted electrons, i.e., those
with sufficient energy to surmount the potential barrier
\begin{equation}
\hat{E}_{c}^{m}(x',x'') = E_{c}^{m}(x',x'') - E_{c}^{0}
\; ,
\label{eq:2}
\end{equation}
where $E_{c}^{m}(x',x'')$ is the maximum value of the potential profile
$E_{c}(x)$ in the interval $[x',x'']$, and $E_{c}^{0}$ is its overall minimum
across the sample. The profile $E_{c}(x)$ comprises the (equilibrium)
conduction band edge potential and the external electrostatic
potential.

From the ballistic electron current $J(x',x'')$, we construct the {\em
thermoballistic current} $J(x)$ at position $x$ inside the sample by summing
up the contributions from all intervals $[x',x'']$ for which $x_{1} \leq x' < x
< x'' \leq x_{2}$, each weighted with the probability of occurrence of the
interval. For simplicity, we here take  this probability in its one-dimensional
form $\exp (-|x'' - x'|/l)$, where $l$ is the mean free path for momentum
relaxation (momentum relaxation length).  We then have
\begin{eqnarray}
&\ & \hspace*{-1.0cm} J(x) = v_{e} N_{c} e^{- \beta E_{c}^{0}} \left\{
w(x_{1},x_{2};l) \; \left[ e^{\beta \mu_{1}} - e^{\beta \mu_{2}} \right] +
\int_{x_{1}}^{x}\frac{dx'}{l} \, w(x',x_{2};l) \, \left[ e^{\beta \mu (x')} -
e^{\beta \mu_{2}} \right] \right.  \nonumber \\ &+& \left.  \int_{x}^{x_{2}}
\frac{dx'}{l} \, w(x_{1},x';l) \, \left[ e^{\beta \mu_{1}} - e^{\beta \mu
(x')} \right] + \int_{x_{1}}^{x} \frac{dx'}{l} \int_{x}^{x_2} \frac{dx''}{l} \,
w(x',x'';l) \, \left[ e^{\beta \mu (x')} - e^{\beta \mu (x'')} \right] \right\}
\; , \nonumber \\
\label{eq:3}
\end{eqnarray}
where
\begin{equation}
w(x',x'';l)  = e^{-|x''-x'|/l} \, e^{-\beta \hat{E}_{c}^{m}(x',x'')} \; .
\label{eq:4}
\end{equation}
The quantities $\mu_{1,2} = \mu(x_{1,2})$ are the chemical potentials on the
{\em contact side} of the interfaces, i.e., immediately {\em outside} of
the sample.  The thermoballistic current $J(x)$ is not conserved by itself;
however, when averaged over the sample, it yields \cite{lip05,lip03} the
physical electron current $J$,
\begin{equation}
\frac{1}{x_{2} - x_{1}}\int_{x_{1}}^{x_{2}} dx \, J(x) =  J   \; .
\label{eq:5}
\end{equation}
Furthermore, the thermoballistic current entering at one end of the sample must
equal the current leaving at the other end \cite{lip05},
\begin{equation}
J(x_{1}^{+}) = J(x_{2}^{-})  \; .
\label{eq:6}
\end{equation}
Substituting expression (\ref{eq:3}) in condition (\ref{eq:5}), we can derive
Volterra-type integral equations for two auxiliary functions $\chi_{1}(x)$ and
$\chi_{2}(x)$, which are distinguished by discontinuities at $x = x_{1}$ and $x
= x_{2}$, respectively.  Using condition (\ref{eq:6}), we determine from
$\chi_{1}(x)$ and $\chi_{2}(x)$ a unique chemical potential $\mu(x)$.  It
exhibits discontinuities at both interfaces, which are related to the Sharvin
interface resistance \cite{sha65}.  From $\mu(x)$, the thermoballistic current
$J(x)$ is calculated via Eq.\ (\ref{eq:3}).  The current-voltage characteristic
is obtained for arbitrary values of $l$ and arbitrary $E_{c}(x)$ in terms of a
``reduced resistance'' $\tilde{\chi}$ composed of $\chi_{1}(x_{2})$ and
$\chi_{2}(x_{1})$.  The thermoballistic density $n(x)$ of electrons making up
the current $J(x)$ is constructed in a similar way.

For electron transport in a homogeneous semiconductor driven by a (constant)
electric field of strength ${\cal E }$, the thermoballistic current $J(x)$ and
density $n(x)$ can be expressed essentially in terms of the dimensionless
parameters $x/S$, $l/S$, and $\epsilon S$, where $\epsilon = \beta e |{\cal
E}|$.  In Fig.\ 2, we show the dependence of $n(x)$ on $x/S$ for $\epsilon S =
1$ and various values of $l/S$.  The initial decrease of $n(x)$ with increasing
$x/S$ is a ballistic effect that reflects the increase of the electron velocity
in the electric field and becomes more pronounced as $l/S$ becomes larger.  The
corresponding thermoballistic current $J(x)$ turns out to differ only
insignificantly from the physical current $J$.  \\

\noindent
{\bf 3.\ Spin-polarized transport in semiconductors}\\[-0.6cm]

We now extend the unified description by allowing spin relaxation to take place
during the electron motion across the ballistic intervals.  The off-equilibrium
spin-polarized current $J_{-}(x) = J_{\uparrow}(x) - J_{\downarrow}(x)$ is
connected with the off-equilibrium spin-polarized density $n_{-}(x) =
n_{\uparrow}(x) - n_{\downarrow}(x)$ through the balance equation
\begin{equation}
\frac{dJ_{-}(x)}{dx} + \frac{n_{-}(x)}{\tau_{s}} = 0 \; ,
\label{eq:7}
\end{equation}
where $\tau_{s}$ is the spin relaxation time. Using this equation in the
ballistic transport regime, in which $J_{-}(x) = 2 v_{e} n_{-}(x)$, we
construct the off-equilibrium ballistic spin-polarized current
$J_{-}(x',x'';x)$ at position $x$ in the ballistic interval $[x',x'']$,
obtaining
\begin{equation}
J_{-}(x', x'';x) = v_{e} N_{c} \, e^{-\beta \hat{E}_{c}^{m} (x',x'')} \left[
A(x') \, e^{- (x-x')/l_{s}} - A(x'') \, e^{- (x''-x)/l_{s}} \right]
\label{eq:8}
\end{equation}
($x_{1} \leq x' < x < x'' \leq x_{2}$), with the ballistic spin relaxation
length $l_{s} = 2 v_{e} \tau_{s}$. The parameter $l_{s}$ comprises
the overall effect of the underlying microscopic spin relaxation mechanisms.
The function $A(x)$ is the ``spin transport function'' defined as
\begin{equation}
A(x') = e^{-\beta [E_{c}^{0} - \mu(x')]} \, \alpha_{-}(x') \; ;
\label{eq:9}
\end{equation}
here, $\alpha_{-}(x') = \alpha_{\uparrow}(x') - \alpha_{\downarrow}(x')$ is the
off-equilibrium ``spin fraction excess'' at the equilibration point $x'$,
which is defined in terms of the spin fractions $\alpha_{\uparrow
\downarrow}(x')$, with $\alpha_{\uparrow}(x') + \alpha_{\downarrow}(x') = 1$.
The spin fractions are related to the spin-resolved thermoballistic chemical
potentials
$\mu_{\uparrow \downarrow}(x')$ via
\begin {equation}
e^{\beta \mu_{\uparrow \downarrow}(x')} = e^{\beta \mu(x')} \,
\alpha_{\uparrow \downarrow}(x') \; .
\label{eq:10}
\end{equation}
Proceeding as in the spinless case, we now sum the (weighted) contributions of
the ballistic spin-polarized current (\ref{eq:8}) over all randomly distributed
intervals $[x',x'']$.  The resulting expression for the off-equilibrium
{\em thermoballistic} spin-polarized current $J_{-}(x)$ is of the form
(\ref{eq:3}), but with the terms in brackets therein replaced with those
obtained by evaluating the bracketed term in Eq.\ (\ref{eq:8}) for the
different cases.  A similar expression is found for the off-equilibrium
{\em thermoballistic} spin-polarized density $n_{-}(x)$.

Substituting the thermoballistic expressions for $J_{-}(x)$ and $n_{-}(x)$ in
Eq.\ (\ref{eq:7}), we arrive at a linear, Fredholm-type integral equation for
the spin transport function $A(x)$,
\begin{eqnarray}
W(x_{1},x;l,l_{s}) \, A_{1} &+& W (x,x_{2};l,l_{s}) \, A_{2} \nonumber
\\ &-& W_{0}(x_{1},x_{2};x;l) \, A(x)  + \int_{x_{1}}^{x_{2}} \frac{dx'}{l} \,
W(x',x;l,l_{s}) \, A(x') = 0 \; ,
\label{eq:12}
\end{eqnarray}
where
\begin{equation}
W(x',x'';l,l_{s}) = w(x',x'';l) \, e^{-|x'' - x'|/l_{s}} \; ,
\label{eq:13}
\end{equation}
\begin{equation}
W_{0}(x_{1},x_{2};x;l) = w(x_{1},x;l) + w(x,x_{2};l) + \int_{x_{1}}^{x_{2}}
\frac{dx'}{l} w(x',x;l) \; ,
\label{eq:14}
\end{equation}
and $A_{1,2} = A(x_{1,2})$.  The solution of the fundamental equation
(\ref{eq:12}) for $x_{1} < x < x_{2}$ determines the spin-polarized electron
transport inside the semiconducting sample, and is obtained in terms of the
values of $A_{1}$ and $A_{2}$ on the contact side of the interfaces at the ends
of the sample.  The latter are given by the spin fraction excesses
$\alpha_{1,2} = \alpha_{-}(x_{1,2})$ and the chemical potentials $\mu_{1,2} =
\mu(x_{1,2})$ in the contacts via Eq.\ (\ref{eq:9}).

For field-driven transport in a homogeneous semiconductor, Eq.\ (\ref{eq:12})
can be converted, by twofold differentiation and elimination of the quantities
$A_{1}$ and $A_{2}$, into a homogeneous integrodifferential equation. Within a
judicious approximation, the latter equation can be reduced to a generalized
spin drift-diffusion equation of the form
\begin{equation}
b_{0}(x) \, \frac{d^{2} A(x)}{dx^{2}} + b_{1}(x) \, \frac{d A(x)}{dx} +
b_{2}(x) \, A(x)  = 0 \; ,
\label{eq:15}
\end{equation}
with coefficient functions $b_{i}(x)$ depending linearly on the function
$\exp[- (\epsilon + 1/l)(x - x_{1})]$.  In the diffusive regime $l/l_{s} \ll
1$, where $b_{0}(x) = 1$, $b_{1}(x) = \epsilon$, $b_{2}(x) = - 1/(l l_{s})$,
Eq.\ (\ref{eq:15}) reduces to the standard spin drift-diffusion equation
\cite{yuf02a} if there the intrinsic spin diffusion length $L$ is identified
with $\sqrt{l l_{s}}$.

With $A(x)$ determined by solving the integral equation (\ref{eq:12}) or, in
special cases, simplified equations like Eq.\ (\ref{eq:15}), we can evaluate
the thermoballistic spin-polarized current $J_{-}(x)$ from the analogue of
expression (\ref{eq:3}).  Dividing by the (total) thermoballistic current
$J(x)$ given by Eq.\ (\ref{eq:3}), we obtain the current spin polarization
\begin{equation}
P_{J}(x) = \frac{J_{-}(x)}{J(x)} \; .
\label{eq:16}
\end{equation}
The density spin polarization is calculated analogously.\\

\noindent
{\bf 4.\ Ferromagnet-semiconductor heterostructures}\\[-0.6cm]

We now consider spin-polarized transport in heterostructures formed of a
homogeneous, nonmagnetic semiconductor and two ferromagnetic contacts, which
are treated as fully degenerate Fermi systems.  Equating the splitting
$\mu_{-}(x) = \mu_{\uparrow}(x) - \mu_{\downarrow}(x)$ of the spin-up and
spin-down chemical potentials in the ferromagnets at $x = x_{1,2}$,
respectively, with that of the semiconductor on the contact side of the
interfaces, we find, using Eq.\ (\ref{eq:10}),
\begin{equation}
[\mu_{-}(x_{1,2})]_{\rm ferromagnet} = [\mu_{-}(x_{1,2})]_{\rm semiconductor} =
\frac{1}{\beta} \ln \left( \frac{1 + \alpha_{1,2}}{1 -\alpha_{1,2}}
\right) \; .
\label{eq:17}
\end{equation}
This condition allows the current spin polarizations $P_{J}(x_{1,2})$ on the
ferromagnet sides of the interfaces to be expressed through the spin fraction
excesses $\alpha_{1,2}$.  Neglecting spin-flip scattering at the interfaces, we
invoke continuity of the polarization at the interfaces. We then obtain
two coupled nonlinear equations from which $\alpha_{1}$ and $\alpha_{2}$, and
hence $J_{-}(x)$ and $P_{J}(x)$, can be calculated in terms of the material
parameters characterizing the ferromagnets and the semiconductor, the bulk
polarizations $P_{1}$ and $P_{2}$ in the left and right ferromagnet,
respectively, and the physical current $J$.  Spin-selective interface
resistances can be introduced via discontinuities of the chemical-potential
splitting $\mu_{-}(x)$ at the interfaces \cite{smi01,ras02a,yuf02a}.

In the case of zero bias, $\epsilon \rightarrow 0$, the solutions of Eq.\
(\ref{eq:15}) are $A(x) \propto \exp( \pm x/L)$, and the position dependence of
the current spin polarization inside the semiconductor is given by
\begin{equation}
P_{J}(x) =  \frac{2 v_{e} N_{c} \bar{l}}{LJ} \, \left[ C_{1} \,
e^{-(x-x_{1})/L} - C_{2} \, e^{-(x_{2}-x)/L} \right]  \; ,
\label{eq:18}
\end{equation}
where $L= \sqrt{\bar{l} l_{s}}$ is the generalized spin diffusion length (or
``polarization decay length''), and $\bar{l} = l l_{s}/(l + l_{s})$.  The
coefficients $C_{1,2}$ are simple functions of the spin fraction excesses
$\alpha_{1,2}$ which, in turn, are given explicitly in terms of the material
parameters, the polarizations $P_{1,2}$, and the current $J$.

In Fig.\ 3, the zero-bias current spin polarization $P_{J}(x)$ for a symmetric
ferromagnet-semiconductor-ferromagnet heterostructure with $S = 1$ $\mu$m at $T
= 300$ K is shown as a function of $x$ for various values of the momentum
relaxation length $l$ and for zero as well as nonzero interface resistances.
For the parameters of the ferromagnets, the values $10^{3}$ $\Omega^{-1}$
cm$^{-1}$ for the bulk conductivities, $60$ nm for the spin diffusion lengths,
and $0.5$ for the bulk polarizations $P_{1,2}$ have been adopted from Ref.\
\cite{yuf02a}.  With a look at recent experiments on the spin dynamics in
n-doped GaAs \cite{kim01}, the values 0.067 $m_{e}$ for the effective electron
mass $m^{\ast}$, $2.0 \times 10^{18}$ cm$^{-3}$ for the equilibrium electron
density, and $1$ $\mu$m for the ballistic spin relaxation length $l_{s}$ have
been chosen for the material parameters of the semiconductor.  We are aware of
the fact that, by considering a specific semiconducting material with fixed
doping concentration, one essentially fixes the value of the momentum
relaxation length $l$.  Therefore, when varying $l$ in a fairly broad range, we
assume the above parameter values (or, at least, their order of magnitude) to
be representative for a class of semiconducting materials that differ in the
strength of the impurity scattering and hence in the magnitude of $l$.

The momentum relaxation length $l$ affects the results shown in Fig.\ 3 in a
twofold way.  (i) It determines the conduction in the semiconductor and thus
the conductivity mismatch with the ferromagnets.  For small values of $l$, this
mismatch is large, leading to a small injected current spin polarization
$P_{J}(0)$.  (ii) It determines the polarization decay length $L$, so that for
small $l$ the polarization dies out rapidly inside the semiconductor.  A
substantial degree of polarization all along the semiconductor is achieved when
the value of $l$ is increased up to a length of the order of the sample length,
in which case the ballistic component becomes prevalent.  Figure 3 also shows
that, by introducing appropriately chosen spin-selective interface resistances,
one may offset the suppression of the injected polarization due to the
conductivity mismatch for small $l$; however, the rapid decay of the
polarization inside the semiconductor cannot be prevented in this way.

For nonzero, constant electric field and $S/L \rightarrow \infty$, i.e.,
disregarding the effect of the right ferromagnet, we have for the current spin
polarization $P_{J}(x_{1})$ injected at the left interface
\begin{equation}
P_{J}(x_{1}) =  \tilde{\chi} \, \Gamma_{J} \, \alpha_{1} \; ,
\label{eq:19}
\end{equation}
where $\tilde{\chi}$ is the reduced resistance entering the current-voltage
characteristic, and the quantity $\Gamma_{J}$ involves the function $A(x)$,
which is obtained by numerically solving Eq.\ (\ref{eq:15}).  Since only the
interface at $x = x_{1}$ enters into consideration, a single nonlinear equation
has to be solved to determine the spin fraction excess $\alpha_{1}$ as a
function of the polarization $P_{1}$.

In Fig.\ 4, we show the dependence of $P_{J}(x_{1})$ on the electric-field
parameter $\epsilon$ for various values of the momentum relaxation length $l$,
the remaining parameter values being the same as in Fig.\ 3. In conformity with
the drift-diffusion results of Ref.\ \cite{yuf02a}, the injected polarization
generally rises with increasing $\epsilon$; however, as in Fig.~3, the main
effect is due to the variation of $l$.  \\

\noindent
{\bf 5.\ Concluding remarks}\\[-0.6cm]

We have outlined the principal ideas of a theory of spin-polarized electron
transport in ferromagnet-semiconductor heterostructures.  It generalizes
previous theoretical treatments based on the drift-diffusion model by
introducing the momentum relaxation length in the semiconductor as a new degree
of freedom, thus allowing a systematic study of the interplay of spin
relaxation and transport mechanism.  By considering illustrative examples, we
have shown that the momentum relaxation length has a significant influence both
on the polarization injected at a ferromagnet-semiconductor interface and on
the decay of the polarization inside the semiconductor.  To study in detail the
influence of the transport mechanism on spin-polarized transport (in
particular, when ballistic effects take over), the identification and design of
classes of novel semiconducting materials is called for.  In this way, new
possibilities to improve the efficiency of spintronic devices may open up.

\newpage

\noindent
{\Large \bf Figure Captions}

\begin{description}

\item[Figure 1:] Schematic diagram showing a semiconducting sample of length
$S$ enclosed between two plane-parallel ferromagnetic contacts.  Illustrated
are expression (\ref{eq:3}) for the thermoballistic current $J(x)$ and its
analogue for the thermoballistic spin-polarized current $J_{-}(x)$.

\item[Figure 2:] The thermoballistic density $n(x)$ inside a homogeneous
semiconductor for constant electric field, plotted versus $x/S$ (assuming
$x_{1} = 0$) for $\epsilon S = 1$ and the indicated values of $l/S$.  The
density is normalized to the constant value it assumes in the diffusive limit
$l/S \rightarrow 0$.

\item[Figure 3:] The zero bias ($\epsilon \rightarrow 0$) current spin
polarization $P_{J}(x)$ (assuming $x_{1} = 0$) along a symmetric
ferromagnet-semi\-conduc\-tor-ferro\-mag\-net structure with $S = 1$ $\mu$m for
the indicated values of the momentum relaxation length $l$.  The solid curves
correspond to zero interface resistance, the dashed curves to interface
resistances of $10^{-7}$ $\Omega$ cm$^{2}$ for spin-up electrons and $2 \times
10^{-7}$ $\Omega$ cm$^{2}$ for spin-down electrons, respectively.  For the
remaining parameter values, see text.

\item[Figure 4:] The injected current spin polarization $P_{J}(x_{1})$ for $S/L
\rightarrow \infty $ as a function of the electric-field parameter $\epsilon$
for the indicated values of the momentum relaxation length $l$.  The solid
curves correspond to zero interface resistance, the dashed curves to interface
resistances of $10^{-7}$ $\Omega$ cm$^{2}$ for spin-up electrons and $2 \times
10^{-7}$ $\Omega$ cm$^{2}$ for spin-down electrons, respectively.  For the
remaining parameter values, see text.

\end{description}

\newpage

\begin{figure}
\vspace*{3.0cm}
\epsfysize 10.0cm
\epsfbox[69 370 506 650]{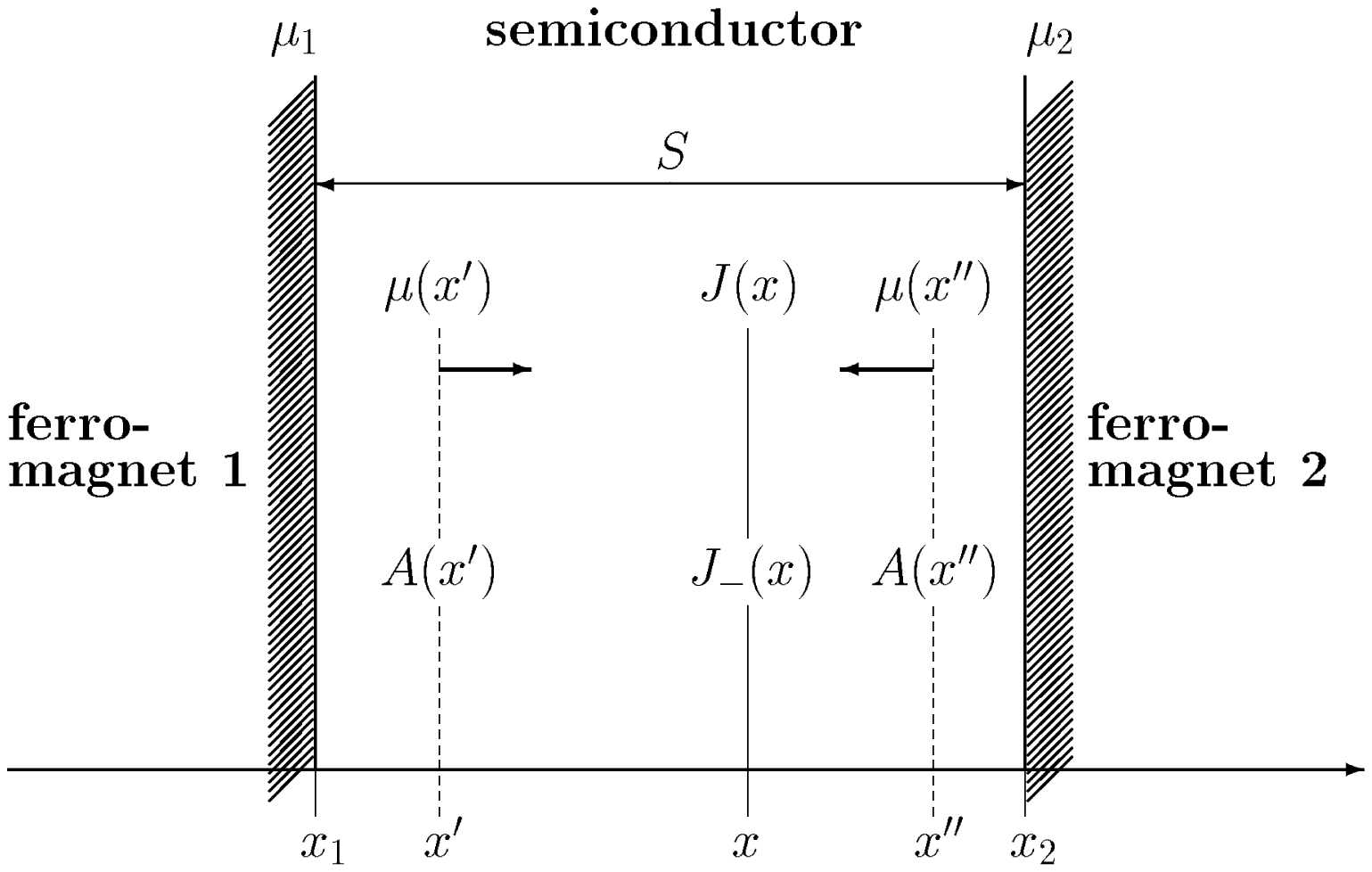}
\vspace*{1.0cm}
\label{fig:1}
\end{figure}
\hspace*{1.0cm}
\vspace{7.0cm}
\begin{center}
{\Large \bf FIGURE 1}
\end{center}

\newpage
\begin{figure}
\vspace*{3.0cm}
\epsfysize 12.0cm
\epsfbox[-25 213 370 623]{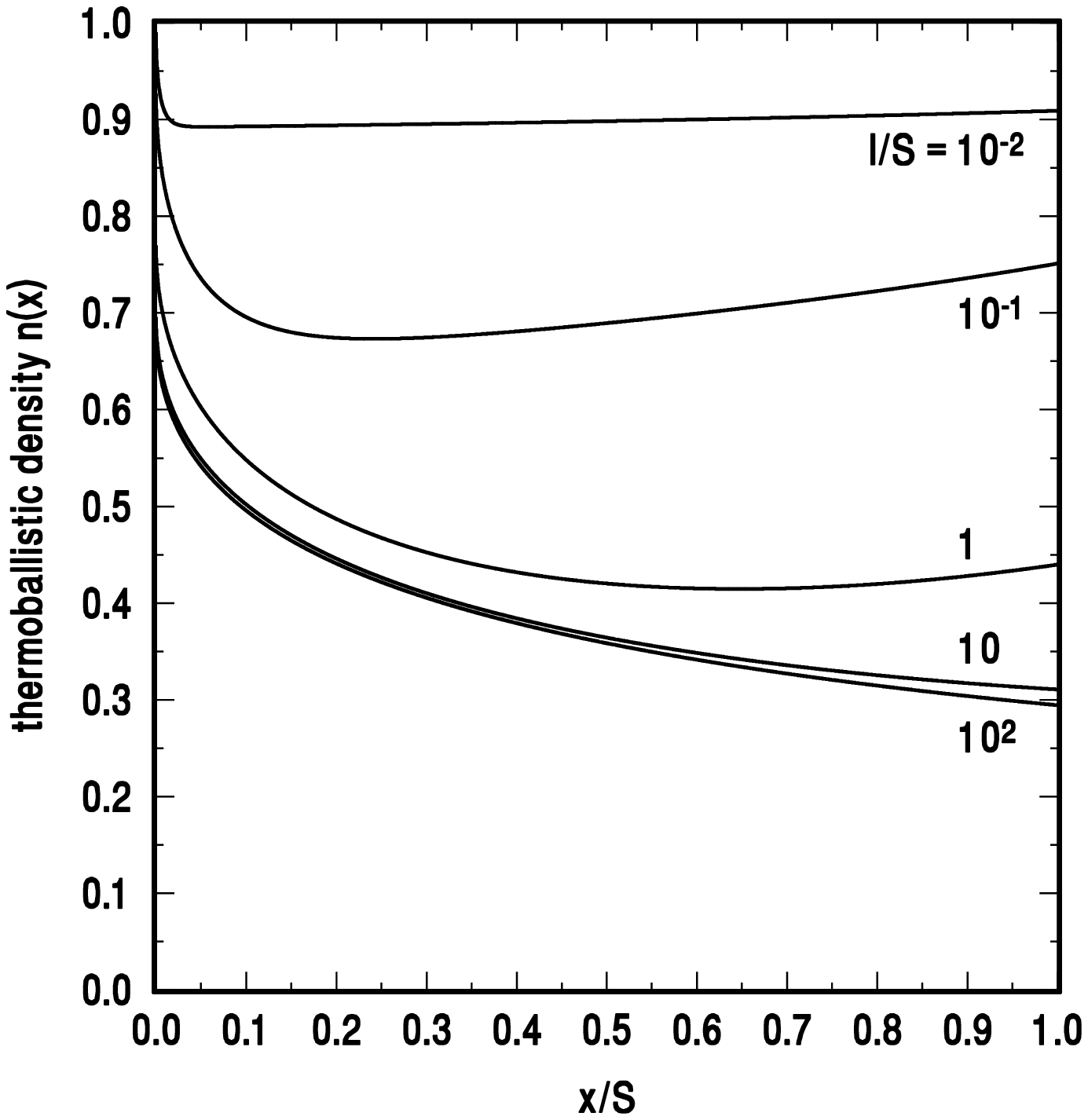}
\vspace*{1.0cm}
\label{fig:2}
\end{figure}
\hspace*{1.0cm}
\vspace{5.0cm}
\begin{center}
{\Large \bf FIGURE 2}
\end{center}

\newpage
\begin{figure}
\vspace*{3.0cm}
\epsfysize 12.0cm
\epsfbox[-25 143 370 553]{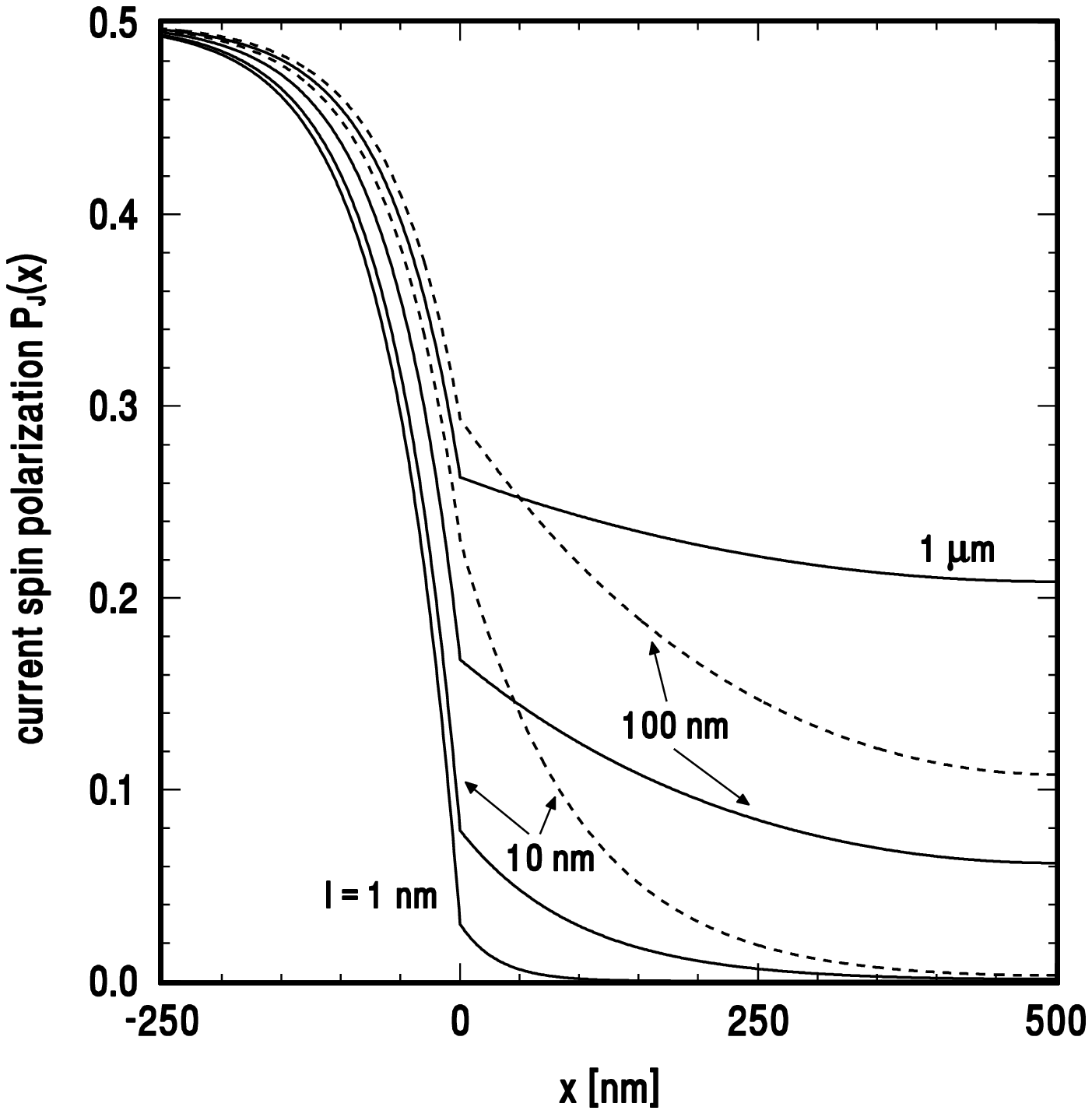}
\vspace*{1.0cm}
\label{fig:3}
\end{figure}
\hspace*{1.0cm}
\vspace{5.0cm}
\begin{center}
{\Large \bf FIGURE 3}
\end{center}

\newpage
\begin{figure}
\vspace*{3.0cm}
\epsfysize 12.0cm
\epsfbox[-25 143 370 553]{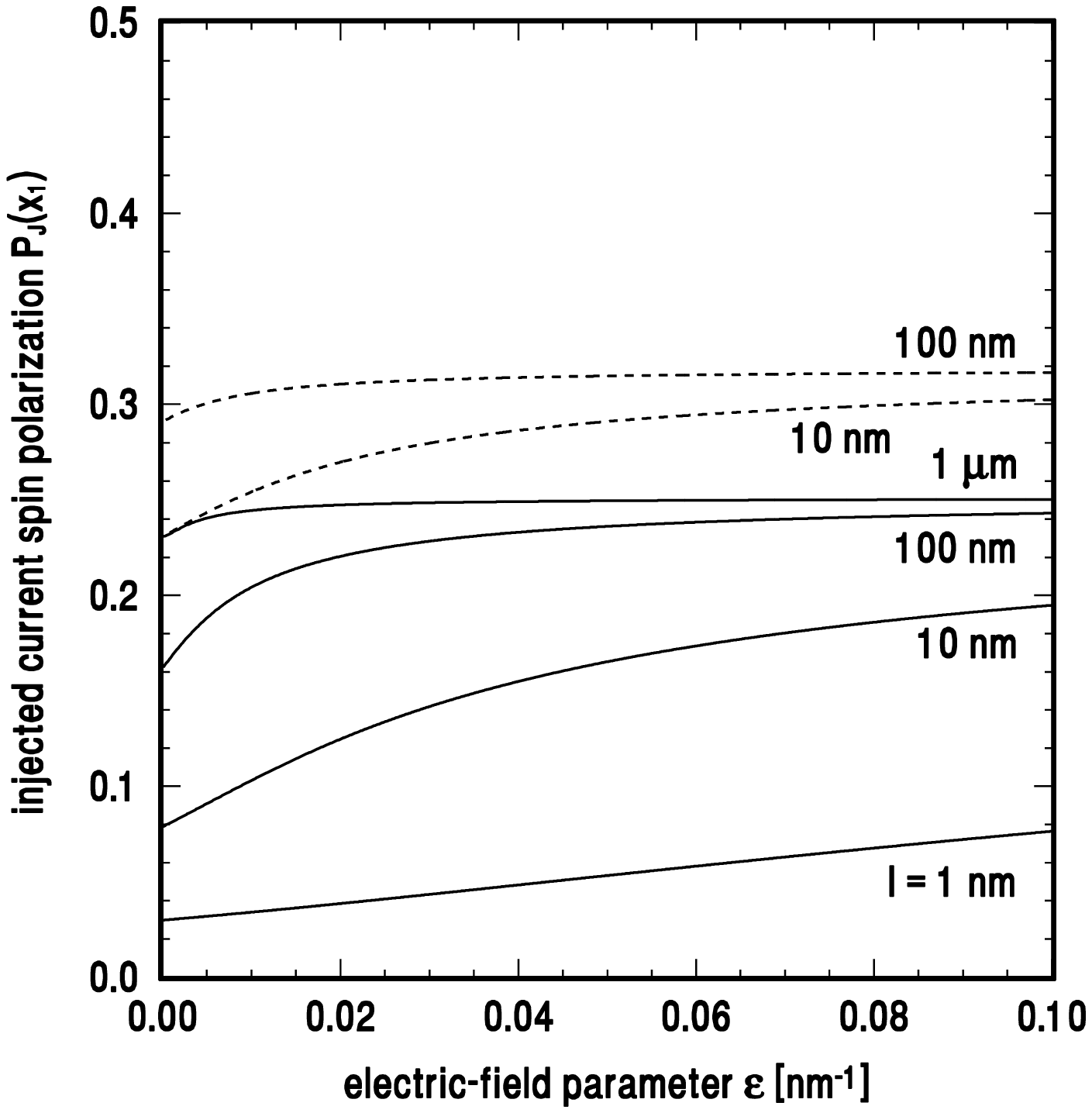}
\vspace*{1.0cm}
\label{fig:4}
\end{figure}
\hspace*{1.0cm}
\vspace{5.0cm}
\begin{center}
{\Large \bf FIGURE 4}
\end{center}

\end{document}